\documentclass[onecolumn,showpacs,preprintnumbers,amsmath,amssymb]{revtex4}

\usepackage{graphicx}
\usepackage{dcolumn}
\usepackage{bm}

\begin{document}

\title{ Bosonization in  SU(N) Gauge Field Theory \\ in Terms of  Phase Transition of Second Kind}

\author{A.V.Koshelkin.}
 \altaffiliation[  $A\underline ~ Kosh@internets.ru$ ]
 {}
\affiliation{Moscow Institute for Physics and Engineering,
Kashirskoye sh., 31, 115409 Moscow, Russia }
\date{\today}

\begin{abstract}
Bosonization of the strong interacting  matter as a process of
arising observable hadrons  is studied in terms of the phase
transition of the second kind. The spectrum of bosons which is
free from the zero point energy is derived . The calculated boson
mass is found to depend self-consistently on both the amplitude of
a gauge field and quark mass. In the framework of the
quasi-classical model\cite{13,14} a hadron mass is calculated in
the case of  bosonization into pions.
\end{abstract}

\pacs{11.15.-q, 03.70.+k, 11.10.-z }

\maketitle

\section{Introduction}

Since observable particles are colorless the process of the
hadronization plays  a key role in QCD. The problem complicates
when hadrons are mesons, since such particles are governed by the
Bose-Einstein statistics while they consist of  interacting
fermions.

Bozonization generally means a description of fermionic systems in
terms of the collective boson degrees of freedom. This concerns
both the quantum field theory\cite{Stone, Fradkin} and condensed
matter\cite{Barci, Neto}. It is preferable to bosonate a fermionic
system by solving  the motion equation for interacting fermions.
Although this way is the most correct and elegant it has been
still made in the case of the Minkowski space-time $(1+1)$.
Following such technique the bosonization of the strong
interaction matter is considered many
times\cite{1,2,3,4,5,6,7,8,9,10,11,12}. The heart of the method
developed in the papers\cite{1,2,3,4,5,6,7} is existence of the
so-called flux tube, when  the strong interacting matter is
suggested to  be in the condition of the longitudinal dominance
and transverse confinement in the Minkowskii space-time. In such
consideration the oscillations of color density are governed by
the Klein-Gordon equation which  mass term contains a boson mass
in an explicit form\cite{1,2,3,4,5,6,7,8,9}. Various mechanisms of
the  $QCD4 \to QCD2 \times QCD2$ fragmentation (or the process of
arising a quark-gluon tube) which is required in the framework of
the consideration developed in Ref.\cite{1,2,3,4,5,6,7,8,9}, are
studied in Ref.\cite{5,6,7}. Separating the longitudinal and
transverse motions, the self-consistent set of equations  for
gauge and fermion fields has been derived\cite{5,6,7} in the
$(1+1)$ Minkowskii space-time.

The bosonization is  considered  in terms of the holographic
description of hadrons in string theory\cite{10,11}. Following the
gauge-string duality\cite{12}  boson masses  haves been
calculated\cite{10,11} due to the $(4\leftrightarrow 10)$ duality.

In the present paper  bosonization as a process of arising
observable hadrons is studied in terms of the QCD lagrangian in
the standard $(1+3)$ Minkiwski space-time without any prior
fragmentation. Considering the bosonization as the equilibrium
phase transition of the second kind, the boson spectrum is derived
in the quasi-classical approximation beyond the fluctuation region
of the transition. The obtained spectrum is free from the zero
point energy. When the confinement phase is governed by the
equations of the self-consistent quasi-classical model\cite{13,14}
a boson mass is calculated provided that a quark-gluon plasma is
bosonated into the lightest bosons (pions).

The paper is organized as follows. The second section contains the
general equations governing the confinement phase and the
relations corresponding to the main approximations. Bosonization
as the phase transition of the second kind is considered in
Section III. The  bosonization in the framework of the
quasi-classical self-consistent model\cite{13,14} is studied in
Section IV.   The applicability of  the obtained results for
describing  observable hadrons is discussed in Section V. Appendix
I contains the main equations of the quasi-classical
self-consistent model\cite{13,14}.

\section{General equations }

The gauge invariant action ${\cal A}$ in the SU(N) field theory
is\cite{15,16,17}:

\begin{eqnarray}
&& {\cal A}  = \int d^4 x  \Bigg\{ {1\over 2}  \left[ {\bar \psi}
(x) \gamma^k   ( i \partial_k +  g T_a A^a_k ) \  \psi (x)  -
{\bar \psi} (x) m  \psi (x) \right] - {1\over 2} \left[ {\bar
\psi} (x) \gamma^k (i {\overleftarrow\partial}_k - g T_a A^a_k )
 \ \psi (x) + {\bar \psi} (x) m
 \psi (x) \right] - \nonumber \\
&&~~~~~~~~ {1\over 16 \pi } F^a_{\mu \nu}  \ F_a^{\mu \nu}
\Bigg\},
\end{eqnarray}
where $F^a_{\mu \nu}$ is the tensor of the non-abelian gauge filed
which is given by the expression:

\begin{eqnarray}
&&   F^a_{\mu \nu} =
\partial_\mu A^a_\nu - \partial_\nu A^a_\mu +  g  \  f^a_{\ b c} A^b_\mu   A^c_\nu
\end{eqnarray}

The action (1) generates the energy-momentum tensor

\begin{eqnarray}
&& T^{\mu \nu } = {i\over 2} \left\{ {\bar \Psi} (x) \gamma^{\mu}
 \partial^{\nu} \Psi (x) - {\bar \Psi} (x)  \gamma^{\mu} {\overleftarrow\partial}^{\nu} \Psi (x)
 \right\} + g (J^a (x) )^{\mu} A_a^\nu (x) + {1\over 4 \pi} \left\{ - F^{\mu i}_a
 (x) (F^a)^{\nu}_i  (x) + {{\cal G}^{\mu \nu}\over 4} F^{i k}_a
 (x) F^a_{i k }  (x)
 \right\} \nonumber \\
\end{eqnarray}

and the motion equations:

\begin{widetext}
\begin{eqnarray}
&& \partial_{\mu} F^{\nu \mu}_a (x)  - g \cdot  f_{ab}^{\ \ c}
A_{\mu}^b (x) F_c^{\nu \mu} (x) = - g  {J_a}^{\nu} (x)
\\ \nonumber   \\ && F_a^{\nu \mu} (x) = \partial^{\nu} A_a^{\mu} (x) -
\partial^{\mu} A_a^{\nu} (x) - g\cdot f_{a}^{\ bc}A_b^{\nu} (x) A_c^{\mu} (x),
\\ \nonumber \\
&& {J_a}^{\nu} (x) = \  {\bar \Psi } (x) \gamma^{\nu}  T_a \Psi
(x) ,
\end{eqnarray}
\end{widetext}

In this way, the fermion  fields $\Psi (x), {\bar \Psi (x) }$ are
governed by the Dirac equation:

\begin{widetext}
\begin{eqnarray}
&& \left\{ i \gamma^{\mu} \left( \partial_{\mu} + i g \cdot
A_{\mu}^a (x) T_a \right) - m \right\} \Psi (x) = 0  \\ \nonumber  \\
&& {\bar \Psi} (x) \left\{ i \gamma^{\mu} \left(
{\overleftarrow\partial}_{\mu} - i g \cdot {A^{\ast}}_{\mu}^a (x)
T_a \right) + m \right\}  = 0; \ \ \ \ \ \ \ .
\end{eqnarray}
\end{widetext}

In Eqs.(1) - (8) we introduce the following notations; $m$ is a
fermion mass, $g$ is the coupling constant; $\gamma^{\nu}$ are the
Dirac matrixes, $x\equiv x^\mu = (x^0 ; {\vec x})$ is a vector in
the Minkowski space-time; $\partial_{\mu} = (\partial /\partial t
; \nabla )$; the Roman letters numerate a basis in the space of
the associated representation of the $SU(N)$ group, so that $a,b,c
= 1 \dots N^2 -1$. We use the signature $ diag \left( {\cal
G}^{\mu \nu} \right) = (1; -1; -1; -1)$ for the metric tensor
${\cal G }^{\mu \nu}$. The line  over $\Psi$ mean the Dirac
conjugation. Summing over any pair of the repeated indexes is
implied.

The symbols $T_a$ in Eqs.(1)-(8) are the generators of the $SU(N)$
group which satisfy the commutative relations and normalization
condition:

\begin{widetext}
\begin{eqnarray}
&& \left[ T_a, T_b \right]_-    =  T_a  T_b - T_b T_a =   i
f_{ab}^{\ \ c} T_c ; \ \ \ \ \ \ f_{ab}^{\ \ c}  = - 2\  i   \ Tr
\left( \left[ T_a, T_b \right]_- T_c \right)  \\
&& Tr \ ( T_a \ T_b )   ={1\over 2} \delta_{ab} ;
\end{eqnarray}
\end{widetext}
where $f_{ab}^{\ \ c}$ are the structure constants of the $SU(N)$
group, which are real and anti-symmetrical with respect to the
transposition in any pair of indexes;   $\delta_{ab}$ is the
Kroneker symbol.

We assume that the field $A_{\nu}^a (x)$ depends on coordinates
via some scalar function $\varphi (x)$  in the Minkowski
space-time which is normally named by the eikonal:

\begin{widetext}
\begin{eqnarray}
&&  A^{\nu}_a ( x )  = A^{\nu}_a (\varphi (x)) .
\end{eqnarray}
\end{widetext}

Let  the  axial gauge  be  for the field $A_{\mu}^a (x)$    :

\begin{widetext}
\begin{eqnarray}
&& \partial^{\mu} A_{\mu}^a  = 0 ;  \ \ \ \ \ \ \ \ k^{\mu} {\dot
A}_{\mu}^a  = 0,
\end{eqnarray}
\end{widetext}
where the dot over the letter means  differentiation with respect
to  the introduced variable $\varphi $ while the vector $k^\mu$
is:

\begin{widetext}
\begin{eqnarray}
&&  k^{\mu}   = \partial^\mu ~ \varphi (x)
\end{eqnarray}
\end{widetext}

The introduced vector$ k^{\mu} $ indicates  the direction along
the eikonal, while Eq.(12) means the local transversion of the
gauge field $ A_{\mu}^a$.

Due to Eq.(11) åhe hamiltonian generated by the action (1) does
not depend explicitly  on the time variably. This means that some
stationary states of fermions,  which energy is $\varepsilon
({\vec p})$,  exist.

Then, the hamiltonian  of interacting fermions can be written as
follows:

\begin{widetext}
\begin{eqnarray}
&& {\cal H} = \int d^3 {\vec x} \ T^{00} = \sum_{{\vec p} ; \sigma
, \alpha} \varepsilon ({\vec p} ) \left[ n_{\sigma , \alpha}
({\vec p}) + (1 - {\bar n}_{\sigma , \alpha} ({\vec p}) ) \right]
+ \int d^3 {\vec x} \ T_g^{00} ,
\end{eqnarray}
\end{widetext}
where $T_g^{00}$ is the zeroth component of the momentum-energy
tensor of the gauge field;  ${ n}_{\sigma , \alpha} ({\vec p}) )$
and ${\bar n}_{\sigma , \alpha} ({\vec p}) )$ are the occupation
numbers of fermions and anti-fermions, respectively.

\section{Bosonization}

In studying   bosonization we follow the assumption that
bosonization starts  when  the  fermion vacuum is full such that
the occupation number of both particles  $n_{\sigma , \alpha}
({\vec p})$ and anti-particles ${\bar n}_{\sigma , \alpha} ({\vec
p})$ are equal to unit:

\begin{widetext}
\begin{eqnarray}
&&   n_{\sigma , \alpha} ({\vec p}) = 1; \ \ \ \ {\bar n}_{\sigma
, \alpha} ({\vec p}) = 1.
\end{eqnarray}
\end{widetext}

As for the a gluon field, we assume that  the number of gluons is
large $n_g \gg 1$ due to the self-interaction of them.  Since $n_g
\gg 1$, the gluon field can be considered quasi-classically.

 Besides that we suggest that fermions and gauge field are in equilibrium.

On the other hand, $A_{\mu}^a$ is self-interacting field that
leads to the generation  of  new quanta of $A_{\mu}^a$ in spite of
the equilibrium state. Then, the creation of additional quanta on
the background of the fullness of the fermion vacuums has to
result in arising new particles  since the  entropy is in maximum.

Let us  consider the matter consisting  of interacting quarks and
gluons. We assume that the matter transits to the deconfinement
phase so that bosons only arise  as observable particles.

We present the gauge field  as a sum of two orthogonal
terms\cite{18,19,20} in the group space such that $A_{\mu}^a$ has
the following form:

\begin{widetext}
\begin{eqnarray}
&& A_{\mu}^a (\varphi )= {\cal A}_{\mu}^a  + e_{\mu}^a \Phi
(\varphi ) ; \ \ \ {\cal A}_{\mu}^a \ e^{\mu}_b = 0  ; \ \ \
e_{\mu}^a \ e^{\mu}_b  = - \delta^a_b ,
\end{eqnarray}
\end{widetext}
where ${\cal A}_{\mu}^a$ is amplitude of the gauge field just
before the phase transition. The amplitude ${\cal A}_{\mu}^a$ is
taken   to be constant in the Minkowski space-time, while $\Phi
(\varphi )$ is a scalar function therein. The field  $\Phi
(\varphi )$ is not to equal to zero in the deconfinement phase,
and plays a role of the order parameter. We note that the
presentation of ${\cal A}_\mu^a $ in the form given by Eq.(17)
means that the phase transition is considered  beyond the
fluctuation region\cite{21}.

Then, the gluon part of the momentum-energy tensor $T_g^{00} $,
which is given by  Eq.(14), is of  the form:

\begin{widetext}
\begin{eqnarray}
&& T_g^{00} = {1\over 16 \pi} \Bigg\{ 4 ( N^2 - 1 ) \left(
\partial^0 \Phi \right)^2 - 2 ( N^2 - 1 ) \left(
\partial^\nu  \Phi \right) \left(
\partial_\nu  \Phi \right) + 2 N g^2 A^2 \Phi^2 + g^2
 f_{a}^{\ b c} \ f^{a}_{\ b_1 c_1 } {\cal A}^{\nu}_b {\cal
A}^{\mu}_c \ {\cal A}_{\nu}^{b_1} \ {\cal A}_{\mu}^{c_1} +  \nonumber \\
&& g^2 N ( N^2 - 1 ) \Phi^4 \Bigg\} ; ~~~~~ - A^2 \equiv {\cal
A}_{\mu}^{a}{\cal A}^{\mu}_{a}
\end{eqnarray}
\end{widetext}

We study the situation when the density $n_0$ of the particles
governed by the field $\Phi$ is not too large, so that

\begin{widetext}
\begin{eqnarray}
 n^{1/3}_0 \lambda_C  \ll 1 ,
\end{eqnarray}
\end{widetext}
where $\lambda_C = 1/M$ is  the Compton wave length of a particle,
which mass is $M$. Such  inequality corresponds to studying the
phase transition beyond the fluctuation region\cite{25}.

Then, the last  term in Eq.(18) is small\footnote{  The detailed
consideration  of this fact is given in the section Discussion}.
As a result, taking into account Eq.(15), we rewrite the
hamiltonian given by Eq.(14) by the following way:

\begin{widetext}
\begin{eqnarray}
&&  {\cal H} =   \sum_{{\vec p} ; \sigma , \alpha} \varepsilon
({\vec p} ) +   {1\over 16 \pi} \int d^3 {\vec x} \Bigg\{ 4 ( N^2
- 1 ) \left(
\partial^0 \Phi \right)^2 - 2 ( N^2 - 1 ) \left(
\partial^\nu  \Phi \right) \left(
\partial_\nu  \Phi \right) + 2 N g^2  A^2 \Phi^2 + \nonumber \\
&&  g^2  f_{a}^{\ b c} \ f^{a}_{\ b_1 c_1 } {\cal A}^{\nu}_b {\cal
A}^{\mu}_c \ {\cal A}_{\nu}^{b_1} \ {\cal A}_{\mu}^{c_1}  \Bigg\}.
\end{eqnarray}
\end{widetext}
We should note here that the hamiltonian (20) is independent on
the color variables in the explicit form.

By changing $\Phi \to {\vec \Phi}$, the relations (20) is easy
generalized to the case when the field ${\vec \Phi} $ is the
triplet of pseudoscalar mesons, where $ {\vec \Phi}$ is the vector
in the isospace.

We expand $ {\vec \Phi} (\varphi) $ over the whole set of the
plane waves:

\begin{widetext}
\begin{eqnarray}
&&  {\vec \Phi} (\varphi) =  \sum_{{\vec q} }  \sqrt{{8\pi \over V
(N^2 - 1)\  \omega ({\vec q})}} \left\{ {\vec e} \  c({\vec q})
\exp(-i q x) + {\vec e}^{\ \ast} \ c^{\dag} ({\vec q}) \exp( i q
x)
 \right\}, \nonumber \\
 && \omega ({\vec q})
 = \sqrt{{\vec q}^{\ 2} + M^2} , \ M^2 = { N g^2  A^2 \over (N^2 - 1)};
    \ \ {\vec e} \ {\vec e}^{\ \ast} = 1 ,
\end{eqnarray}
\end{widetext}
where ${\vec e}$ is the unit vector in the isospace; $c({\vec q})$
and $c^\dag ({\vec q})$ are the operators of annihilation and
creation of the on-shell  particle ($ q^2 = M^2  $) with the
4-momentum $q = (\omega ({\vec q}) ; {\vec q})$. The operators
$c({\vec q})$ and $c^\dag ({\vec q})$ satisfy the standard
Bose-Einstein commutative relations.

Let us substitute the expansion given by Eq.(21) into the formula
(20) and average the obtained relation over the vacuum of the
field ${\vec \Phi} $. As a result,   we derive the energy of the
particles governed by the pseudoscalar field ${\vec \Phi} $:

\begin{widetext}
\begin{eqnarray}
&&  E =   \sum_{\vec q}  \omega ({\vec q} ) \left(  < c^{\dag}
({\vec q}) \ c({\vec q}) > \right) + \left\{{1\over 2} \sum_{\vec
q} \omega ({\vec q} ) +
    \sum_{{\vec
p} ; \sigma , \alpha} \varepsilon ({\vec p} ) +  g^2  f_{a}^{\ b
c} \ f^{a}_{\ b_1 c_1 }  {\cal A}^{\nu}_b {\cal A}^{\mu}_c \ {\cal
A}_{\nu}^{b_1} \ {\cal A}_{\mu}^{c_1}
 \right\},
\end{eqnarray}
\end{widetext}
where the angle brackets mean averaging over the pseudoscalar
vacuum.

Since the vacuum of  arising pseudoscalar particles should be
empty the term in the curl brackets has to be equal to zero. This
condition determines the spectrum $\varepsilon ({\vec p})$ of
quarks via the gauge field ${\cal A}_{\mu}^{c}$ just before the
phase transition. We should note here that  the last term in the
curl bracket should be negative.

As a result, we obtain  the  energy spectrum of scalar  hadrons:

\begin{widetext}
\begin{eqnarray}
&&  E_h =   \sum_{\vec q}  \omega ({\vec q} ) N_h ({\vec q})   ; \
\ \ \omega ({\vec q})
 = \sqrt{{\vec q}^{\ 2} + M^2}
\end{eqnarray}
\end{widetext}

\noindent provided that

\begin{widetext}
\begin{eqnarray}
&& \left\{{1\over 2} \sum_{\vec q} \omega ({\vec q} ) +
    \sum_{{\vec
p} ; \sigma , \alpha} \varepsilon ({\vec p} ) +  g^2  f_{a}^{\ b
c} \ f^{a}_{\ b_1 c_1 }  {\cal A}^{\nu}_b {\cal A}^{\mu}_c \ {\cal
A}_{\nu}^{b_1} \ {\cal A}_{\mu}^{c_1}
 \right\} = 0,
\end{eqnarray}
\end{widetext}
where $N_h ({\vec q})$ is the number of the on-shell hadrons with
the 4-momentum  $q = (\omega ({\vec q}) ; {\vec q})$.

\section { Bosonization in quasi-classical model}

Let us apply the results obtained in the previous sections to the
calculation of a boson mass in terms of the self-consistent
quasi-classical model developed in Ref.\cite{13,14} (see, also
Appendix):

In this case the convolution in Eq.(22)  is equal to\cite{14}

\begin{widetext}
\begin{eqnarray}
&& -  f_{a}^{\ b c} \ f^{a}_{\ b_1 c_1 }  {\cal A}^{\nu}_b {\cal
A}^{\mu}_c \ {\cal A}_{\nu}^{b_1} \ {\cal A}_{\mu}^{c_1}  =  (N^2
-1)\sum\limits_{\sigma \alpha }\int {d^3 p \over p^{(0)} (2\pi)^3
}.
\end{eqnarray}
\end{widetext}

Then, the boson mass $M$ is given by a formula:

\begin{widetext}
\begin{eqnarray}
&& M^2 \thickapprox {2 ~N \ N_f ~\alpha_s \over 2 \vert C \vert
\pi } Q^2;
\nonumber \\
&& \alpha_s = {g^2 \over 4 \pi}, \ \ \  C  = -  f_{a}^{\ b c} \
f^{a}_{\ b_1 c_1 }  \cos (\varphi_b - \varphi_{b_1} ) \cos
(\varphi_c - \varphi_{c_1} ) > 0,
\end{eqnarray}
\end{widetext}
where $\alpha_s$ is the strong interaction coupling constant, $Q$
is the transferred momentum corresponding to the
confinement-deconfinement phase transition which is of the order
of the phase transition temperature . The parameters
$\varphi_{b,c,b_1 , c_1 }$ are the phases of the amplitudes ${\cal
A}^\nu_{b,c,b_1 , c_1 }$ which are fixed such that the convolution
$C$ is  negative.

In the case  $N_f = 2; \ N=3$, we have:

\begin{widetext}
\begin{eqnarray}
&& M \thickapprox \sqrt{{12 \alpha_s \over \pi }}  Q .
\end{eqnarray}
\end{widetext}

The last formula establish relation  of the hadron mass $M$ to the
momentum of interacting particles in the matter which depends
strongly on the matter temperature $T$.

If we set $T =  Q = 213 MeV$\cite{22}, then $\alpha_s = 0.12$. As
a result we obtain:

\begin{widetext}
\begin{eqnarray}
 &&   M\approx 144 MeV \ \ \ ; \vert C \vert \sim 1 ,
\end{eqnarray}
\end{widetext}
that corresponds to the pion mass.

Although  the derived pion mass is very nearly to the tabulated
date Eq.(26)  should be mainly treated as the formula giving the
relation of a hadron mass to the temperature of phase transition.
Particular, when the phase transition temperature is around 200
MeV the result for the mass of observable particles is found to be
correct.

In the case  $\varepsilon ({\vec p}) = \sqrt{{\vec p}^{\ 2} + m^2
}$\cite{13,14}, the condition (24) leads to

\begin{widetext}
\begin{eqnarray}
&& {V_0 \over V } \sim \alpha_s \ll 1,
\end{eqnarray}
\end{widetext}
where $m$ is a quark mass; $V_0$ and $V$ are  the volumes occupied
by the quark-gluon and hadron phases, respectively. The derived
inequality is  expectable  and means that  the volume occupied by
hadrons is much greater as compared with one for a quark-gluon
plasma.

\section{Discussion }

We discuss the obtained results in terms of the key assumption
consisting in  negligibility of the last term in (18). To do it we
calculate the contribution of this term , $\Delta E$, into the
spectrum (23) when the field ${\vec \Phi}$ is given by Eq.(21).

In this case direct calculations gives:

\begin{widetext}
\begin{eqnarray}
&& \Delta E \backsimeq \frac{1}{16\pi}~ \int d^3 {\vec x }
\Big\langle g^2 N (N^2 - 1) \Phi^4 \Big\rangle = \frac{6}{16\pi} ~
g^2 N (N^2 - 1) ~ \int \frac{ V^2~d^3 {\vec q}}{(2\pi )^3}
\left(\frac{8\pi}{V~(N^2 -1) \omega ({\vec q}) } \right)^2 N_h^2
({\vec q }) ,
\end{eqnarray}
\end{widetext}
where  $ N_h^2 ({\vec q })$ is the mean value of the occupation
number of hadrons; $\omega ({\vec q}) $ is the hadron energy given
by Eq.(20), $V$ is the volume occupied by the hadrons. The factor
$6$ has arisen due to taking into account the transpositions in
the operators $c^\dag$ and $c$. The angle brackets mean averaging
over the  hadron vacuum.

When  free hadrons are  in  equilibrium the number of them, $N_h$,
is governed by the Bose-Einstein distribution function with the
zeroth potential, $\mu = 0$ . Then, we derive from the last
formula:

\begin{widetext}
\begin{eqnarray}
&& \Delta E = \frac{48 ~\alpha_s ~ N M }{(N^2 -1)  } ~
\int\limits_0^\infty \frac{ \xi^2~d\xi }{\xi^2 + 1 } ~~ \frac{1}{
\left( e^{\frac{\sqrt{\xi^2 +1}}{T/M_h}} -1 \right)^2 }
 , \ \ \ \  \alpha_s = \frac{g^2}{4\pi}.
\end{eqnarray}
\end{widetext}

Calculation of the energy according to the formula (23) results
in:

\begin{widetext}
\begin{eqnarray}
&&  E = \frac{V   M^4 }{2 \pi^2 } ~ \int\limits_0^\infty \frac{
(\xi^2 + 1 )^{1/2}~ \xi^2~d\xi }{  \left( e^{\frac{\sqrt{\xi^2
+1}}{T/M_h}} -1 \right) }.
\end{eqnarray}
\end{widetext}

As a result, the correction to the energy (23) due to the
$\Phi^4$-term is:

\begin{widetext}
\begin{eqnarray}
&& \frac{\Delta E}{E} = \frac{96 \pi^2~ N  ~ \alpha_s}{(N^2 -1) }
~ \frac {\int\limits_0^\infty (\xi^2 + 1 )^{-1}~ \xi^2~   \left(
e^{\frac{\sqrt{\xi^2 +1}}{T/M_h}} -1 \right)^{-3} ~ d\xi }
{\int\limits_0^\infty  (\xi^2 + 1 )^{1/2}~ \xi^2~   \left(
e^{\frac{\sqrt{\xi^2 +1}}{T/M_h}} -1 \right)^{-1} ~ d\xi } ~
\left( \frac{1}{V M^3} \right) \sim n_0 \lambda_C^3 ; ~~~~~
\lambda_C = M^{-1}.
\end{eqnarray}
\end{widetext}

It follows form the last formulae that the correction is
proportional to the gas parameter $n_0 ~ \lambda_C^3 \ll 1 $ which
has been already introduced by Eq.(19).

In the RHIC and SPS experiments\cite{22,23} the characteristic
temperature of hadronic phase is $T_h \thickapprox 200 MeV $ while
the radius of the fireball is $r_F \ge 10 F$. Then,  calculating
the integrals in Eqs.(31), (33) numerically, we derive:

\begin{widetext}
\begin{eqnarray}
&& \frac{\Delta E}{E} \lesssim 1 \cdot 10^{-3},
\end{eqnarray}
\end{widetext}
that proves  reasonability of the used approximation.

The estimations of the corrections to the  spectral distribution
of the energy $\delta (\Delta E )/ \delta N_h $ can be directly
derived form Eqs.(30), (31), and result in:

\begin{widetext}
\begin{eqnarray}
&& \frac{\delta (\Delta E)}{\omega (\vec q) ~ \delta N_h ({\vec
q})} = \frac{192\pi^2 ~ N ~ \alpha_s}{(N^2 -1) } ~ \frac {1}{(q^2
/M^2 + 1 )^{3/2}} ~ \left( \frac{1}{V M^3} \right)   \left(
e^{\frac{\sqrt{q^2 /M^2 +1}}{T/M_h}} -1 \right)^{-1}
 \sim n_0 \lambda_C^3 ; ~~~~~
\lambda_C = M^{-1}.
\end{eqnarray}
\end{widetext}

The numerical calculations according to the formula (35) gives
that the maximum of the value of $\frac{\delta (\Delta E)}{\omega
(\vec q) ~ \delta N_h ({\vec q})}$  (when $q=0$)
 is no more than $0.24$.

The carried out estimation have been made in the case of  the
hadronization at the temperature $T_c = 200 Mev $ which  is likely
to be upper magnitude for $T_c $. In the assumption of the
adiabatic model of the expending matter\cite{24}, when

\begin{widetext}
\begin{eqnarray}
&& V ~ T^3 = ~const,
\end{eqnarray}
\end{widetext}
the both Eqs. (33),(35) decrease   with temperature decreasing.
This means that correction to the energy spectrum due to $\Phi^4$
terms become smaller if the realistic hadronization temperature
appears to be  less than the considered $T_c = 200 MeV$.

We should point out that  the inequality $n^{1/3}_0 \lambda_C \ll
1 $ and   Eq.(29) are not in contradiction. It follows from the
relations:

\begin{widetext}
\begin{eqnarray}
&& n_0 \lambda^3_C = {n_0 \over M^3 } \sim {N_h  \over M^3  V}
\sim {V_0 \over V } \ll 1,
\end{eqnarray}
\end{widetext}
where $N_h$ is the number of hadrons which mass is $M$.

The formula (29) can be also  treated in terms of the equilibrium
phase transition. Since the transition is  equilibrium one the
entropy is in a maximum. In order to the entropy keeps its maximum
, when the phase volume of the confinement phase is decreased due
to the bosonization, the phase volume of the hadronic phase has to
be increased.

We should note here that that the relation (26), (28) can be also
considered as a way to  calculate the strong coupling constant
$\alpha_s$. Provided that a bosom mass has been already known due
to experiments, Eqs.(25) and (27) allow  us to estimate the value
of the constant $\alpha_s$   in various energy regions.

\section {Conclusion}

The bosonization as the phase transition of the second kind is
considered in terms of the QCD gauge invariant lagrangian  in the
standard $(1+3)$ Minkiwski space-time. In the quasi-classical
approximation the spectrum of bosons which is free from the zero
point energy is derived beyond the fluctuation region of the
transition\cite{21}. When the confinement phase is governed by the
equations of the self-consistent quasi-classical model\cite{13,14}
the boson mass is calculated provided that the bosonization into
the lightest bosons (pions) only takes place. The obtained mass
are found to correspond quantitatively to the pion mass provided
that the phase transition temperature  is near $200 MeV$.

\appendix
\section{}

The equations (4)-(8) have the self-consistent solutions which can
be written as follow.\cite{13,14} The fermion field is governed by
the formula:

\begin{widetext}
\begin{eqnarray}
&& \Psi (x) = \sum\limits_{\sigma ,  \alpha } \int {d^3 p \over
\sqrt{2p^0 \ } (2\pi)^3 } \left\{ {\hat a}_{\sigma , \alpha}
({\vec p}) \Psi_{\sigma , \alpha }  ( x, p ) +
{\hat b}^\dag_{\sigma , \alpha } ({\vec p}) \Psi_{ - \sigma , \alpha }  ( x, - p )  \right\} \nonumber \\
&& {\bar \Psi (x)}  = \sum\limits_{\sigma ,  \alpha} \int {d^3 p
\over \sqrt{2p^0 \ } (2\pi)^3 } \left\{ {\hat a}^\dag_{\sigma ,
n\alpha} ({\vec p})\ {\bar \Psi}_{\sigma , \alpha }  ( x, p ) +
{\hat b}_{\sigma , \alpha} ({\vec p}) \ {\bar \Psi}_{ - \sigma ,
\alpha } ( x,
 - p ) \right\} ,
\end{eqnarray}
\end{widetext}
where the symbols ${\hat a}^\dag_{\sigma , \alpha} ({\vec p})
;{\hat b}^\dag_{\sigma , \alpha} ({\vec p})$ and ${\hat a}_{\sigma
, \alpha} ({\vec p}); {\hat b}_{\sigma , \alpha} ({\vec p})$ are
the operators of creation and cancellation of a fermion (${\hat
a}_{\sigma , \alpha} ({\vec p}) ;{\hat a}^\dag_{\sigma , \alpha}
({\vec p})$) and anti-fermion (${\hat b}_{\sigma , \alpha} ({\vec
p}) ;{\hat b}^\dag_{\sigma , \alpha} ({\vec p})$) , respectively.
In this way,  ${\hat a}_{\sigma , \alpha} ({\vec p})$ and $ {\hat
a}^\dag_{\sigma , \alpha} ({\vec p})$; ${\hat b}_{\sigma , \alpha}
({\vec p})$ and $ {\hat b}^\dag_{\sigma , \alpha} ({\vec p})$
satisfy the standard commutative relations for the fermion
operators.

The function $\Phi_{\sigma, \alpha} (x, p)$ has a form:

\begin{widetext}
\begin{eqnarray}
&& \Psi_{\sigma, \alpha} (x, p) =  \Phi_{\sigma, \alpha} (x, p) =
\ \cos \theta \cdot \exp \left(  - i g^2 {( N^2 - 1) A^2 \over  2
N ( p k ) } \varphi - i p x \right)\  \Bigg \{ \left( 1 - i g T_a
{ \tan \theta \over \theta ( p k ) }  \ \int\limits_0^\varphi d
\varphi^\prime \left( A_\mu^a p^\mu \right) \right) + \nonumber
\\
&& {g \left( \gamma^\nu k_\nu \right) \left( \gamma^\mu A_\mu^a
\right) \over 2 ( p k ) } \cdot \Bigg[ {\tan \theta \over \theta}\
T_a  +  {g\over  ( p k) } \ {1\over 2N} \  \left( - i {\tan \theta
\over \theta} + {g\over  ( p k)}  {\theta - \tan \theta \over
\theta^3} T_b \ \int\limits_0^\varphi d \varphi^\prime \left(
A_\mu^b p^\mu \right) \right) \int\limits_0^\varphi d
\varphi^\prime \left( A_\nu^a p^\nu \right) \Bigg] \Bigg\}\
u_\sigma (p) \cdot
 v_\alpha ;  \nonumber \\ \nonumber \\
 && \theta =  { g \over  ( p k ) } \sqrt{{1\over 2N}} \left(
\int\limits_0^\varphi d \varphi^\prime \left( A_\mu^a (
\varphi^\prime ) \  p^\mu \right) \ \int\limits_0^\varphi d
\varphi^{\prime \prime}\left( A_a^\mu ( \varphi^{\prime \prime })
\ p_\mu \right) \right)^{1\over 2} ; \ \ \ \  (\partial_\nu\ k^\nu
) = (\partial_\nu\
\partial^\nu ) \varphi (x) = 0 .
\end{eqnarray}
\end{widetext}

In this way, the spinors $u_\sigma (p)$ satisfy the relations:

\begin{widetext}
\begin{eqnarray}
&&  \sigma^{\mu \nu } k_\mu A_\nu (\varphi = 0 )  \ u_\sigma (p) =
0 \ ;  \ \ \ \ \  {\bar u}_\sigma (p) u_\lambda (p^\prime ) = \pm
2m \ \delta_{\sigma \lambda} \ \delta_{p p^\prime} ; \ \ \ p^2 =m
^2 ,
\end{eqnarray}
\end{widetext}
where $u_\sigma (p)$ are the bispinors of the free Dirac field.
The plus and minus signs in Eq.(A3) correspond to the Dirac scalar
production of the spinors $u_\sigma (p)$ and $u_\sigma (- p)$,
respectively, while the function $\Psi_{\sigma, \alpha} (x, p )$
are normalized by the condition:

\begin{widetext}
\begin{eqnarray}
&& \int d^3 x  \Psi_{\sigma, \alpha}^\ast (x, p^\prime )
\Psi_{\sigma, \alpha} (x, p)   = (2\pi )^3 \delta^3 ( {\vec p} -
{\vec p}^{\ \prime} ).
\end{eqnarray}
\end{widetext}

As for the gauge field it is determined by the equations:

\begin{widetext}
\begin{eqnarray}
&&  A^{\nu}_a (\varphi)  =   A  \left(   e_{(1)}^\nu (\varphi)
\cos \left( \varphi (x)  + \varphi_a  \right)  +   \ e_{(2)}^\nu
(\varphi)  \sin \left( \varphi ( x ) + \varphi_a \right) \right) +
{\cal B}_a  \
\partial^\nu
\varphi (x) \, \nonumber \\
&&  e_{(1)}^\nu  {e_{(2)}}_\nu =  e_{(1)}^\nu k_\nu = e_{(2)}^\nu
k_\nu = 0 ; \ \ \ {\dot e_{(1)}^\nu } = e_{(2)}^\nu ; \ \ \  {\dot
e_{(2)}^\nu } = - e_{(1)}^\nu   ; \ \ \ \ \
 k^\nu
\equiv
\partial^\nu \varphi (x) ,
\end{eqnarray}
\end{widetext}
where    $e_{(1), (2)}^\nu (\varphi)$ are the space-like 4-vectors
on the wave surface $\varphi (x)$ which are independent on the
group variable $a$; the symbols $A$, ${\cal B}_a$ and $\varphi_a$
are some constants in the Minkowski space-time. They are
determined via the initial condition of the considered problem.

The fermion and gauge fields are found to not be independent and
to relate one to  another by the formulae:

\begin{widetext}
\begin{eqnarray}
&& 2 f_{ab}^{\ \ c} \sin \left(\varphi_b - \varphi_c \right) =
 f_{ab}^{\ \ c}\left\{ \ f_{c}^{\ \ sr}  \cos \left(
\varphi_b - \varphi_r \right) + \left\{ \cos \left( \varphi_b -
\varphi_r \right) \  \cos \left( \varphi_s - \varphi_a \right)
\right\} {f_{c}^{\ \ bs }  \over  N } \right\} {\cal B}_s  ;
\end{eqnarray}
\end{widetext}

\begin{widetext}
\begin{eqnarray}
&&  A^2 \cdot C    = - (N^2 -1)\sum\limits_{\sigma \alpha }\int
{d^3 p \over p^{(0)} (2\pi)^3 } \langle {\hat a}^\dag_{\sigma ,
\alpha} ({\vec p}) {\hat a}_{\sigma , \alpha} ({\vec p}) + {\hat
b}_{\sigma , \alpha} ({\vec p}) {\hat b}^\dag_{\sigma , \alpha}
({\vec p}) \rangle ,
\end{eqnarray}
\end{widetext}
where
\begin{widetext}
\begin{eqnarray}
&& C =  \  f_{ab}^{\ \ c}\ f_{c}^{\ \ sr} \left\{ \cos \left(
\varphi_b - \varphi_r \right) \  \cos \left( \varphi_s - \varphi_a
\right) \right\} < 0 ,
\end{eqnarray}
\end{widetext}

\begin{widetext}
\begin{eqnarray}
&& \left( \partial_\mu  \varphi (x) \right) \cdot \left(
\partial^\mu  \varphi (x) \right) = 0 ;\ \ \ \  \left( \partial_\mu  \partial^\mu \right) \varphi (x)  =
0.
\end{eqnarray}
\end{widetext}

\end{document}